**A theoretical framework to consider energy transfers within growth theory**

Benjamin Leiva[a,*], Octavio Ramirez[a], John R. Schramski[b]

[a] College of Agriculture and Environmental Science, University of Georgia, Athens, U.S.A.

[b] College of Engineering, University of Georgia, Athens, U.S.A.

[*] Corresponding author

*Email addresses*: bnlc@uga.edu (B. Leiva), oramirez@uga.edu (O. Ramirez), jschrams@uga.edu (J. R. Schramski).

**Abstract**

Growth theory has rarely considered energy despite its invisible hand in all physical systems. We develop a theoretical framework that places energy transfers at centerstage of growth theory based on two principles: (1) goods are material rearrangements and (2) such rearrangements are done by energy transferred by prime movers (e.g. workers, engines). We derive the implications of these principles for an autarkic agent that maximizes utility subject to an energy budget constraint and maximizes energy surplus to relax such constraint. The solution to these problems shows that growth is driven by positive marginal energy surplus of energy goods (e.g. rice, oil), yet materializes through prime mover accumulation. This perspective brings under one framework several results from previous attempts to insert energy within growth theory, reconciles economics with natural sciences, and provides a basis for a general reinterpretation of economics and growth as the interplay between human desires and thermodynamic processes.

**Nomenclature**

| | | | | | |
|---|---|---|---|---|---|
| 1 | $G$ | Total energy expenditure | 12 | $Q_n$ | Quantity of non-energy good $n$ |
| 2 | $I$ | Total energy income | 13 | $\alpha_e$ | Mg. energy surplus of energy good $e$ |
| 3 | $x_l$ | Prime mover of type $l$ | 14 | $\phi_l$ | Mg. energy surplus of prime mover $l$ |
| 4 | $p_l$ | Power rate of $l$ | 15 | $E$ | Total energy surplus |
| 5 | $P$ | Aggregate power level | 16 | $E_e$ | Energy surplus secured with good $e$ |
| 6 | $\varepsilon_l$ | Direct energy transferred by $l$ | 17 | PES | Primary Energy Sources |
| 7 | $\omega_l$ | Total energy transferred by $l$ | 18 | $\varphi$ | Proportion of useless energy surplus |
| 8 | $\delta_e$ | Energy content of energy good | 19 | $\lambda$ | Mg. utility of energy surplus |
| 9 | $\gamma_e$ | Marginal embodied energy of good $e$ | 20 | $\zeta$ | Energy good equilibrium |
| 10 | $\gamma_e^A$ | Average embodied energy of good $e$ | 21 | $\eta_n$ | Quantity elasticity of $\gamma_n^A$ |
| 11 | $Q_e$ | Quantity of energy good $e$ | 22 | $r_l$ | $x_l$'s maximum rate of accumulation |

Energy has rarely been part of the narratives developed by economists to study economic growth (Aghion & Howitt, 2009; Beinhocker, 2006; Galor, 2011; Galor & Weil, 2000; Lucas, 1990; Romer, 1990; Solow, 1956). Although this omission has been justified on energy's low cost share in production (Perry, 1977; Denison, 1985; Kümmel et al., 2015), it is surprising given that growth theory traditionally began with descriptions of the material conditions of production (Perrings, 1987), and energy's invisible hand in all physical systems was unveiled by natural scientists over a century ago (Boltzmann, 1886; Maxwell, 1872; Ostwald, 1892). Such omission also stands against the extensive documentation of the role played by energy — and the systems that use it — in human history (Cottrell, 1955; Gillett, 2006; Herrmann-Pillath, 2015; Lotka, 1925; Odum, 1971; Rees, 2012; Smil, 2016; White, 1943).

Attempts to consider energy within growth theory have stemmed from scholars focused on the physical conditions of an economy and on energy transitions. The former stress that energy is the potential to do work and that production upgrades the organization of matter with free energy (Ayres et al., 2003; Cleveland et al., 1984; Georgescu-Roegen, 1971; Gillett, 2006; Herrmann-Pillath, 2015; Lindenberger & Kümmel, 2011; Warr, et al., 2008). The latter argue that growth theories such as Solow (1956) and Galor & Weil (2000) can include energy to explain major evolutionary transitions, and show that energy has an important role in explaining growth (Court et al., 2018; Fröling, 2011; Kander & Stern, 2014; Tahvonen & Salo, 2001). However, none of them consider that *energy transfers* are the essence of all productive processes.

How can *energy transfers* be placed at center stage of growth theory? The objective of this paper is to do this by articulating marginal analysis —economists' canonical methodology— with key insights from physics. Our model highlights the relevance of consumers' energy budget constraint, the centrality of prime movers in doing energy transfers, and the existence of marginal embodied energy curves for all goods. Also, this approach brings under one general framework an array of results derived by previous attempts to insert energy within growth theory, like the importance of energy in growth and the constraint that the energy surplus secured by energy sectors imposes on the existence of non-energy sectors. Lastly, by informing economists' canonical modelling technique with some of physics' key concepts, the paper contributes to the dialogue between economics and natural science.

## 1. The energy budget constraint

We first present the energy budget constraint (equation (3)), which is our point of departure from traditional theory. The rationale behind this constraint lays on two principles: (1) goods are material rearrangements of raw materials that satisfy an agent's desires, and (2) material rearrangements are done by energy transfers. The former principle is a natural implication of Lavoisier's Law, as the impossibility of creating matter implies that production can only create order. The latter principle is supported by the broad consensus among natural scientists that material change requires energy transfers.[1]

Under these principles, production is the process of transferring energy to rearrange raw materials into final goods, and therefore depends primarily on the two main components of energy transfers. The first are the energy goods from where energy is transferred. These have energy content (i.e. joules per gram) and must be compatible with one or more prime movers, yet can be as diverse as rice, oil, and electricity. The second are the prime movers that trigger the transfers. These have power rates (i.e. watts) and must be compatible with one or more energy goods, yet can be as diverse as peasants, engineers, diesel engines, and computers.

The indivisible centrality of these two components for production (Cottrell, 1955; Debeir et al., 1991; Weissenbacher, 2009; White, 1943) highlights the conceptual shortcoming of modelling output as a function of labor, capital, and energy (e.g. Berndt & Wood (1975)). The services provided by labor and machinery during production are the energy they transfer to rearrange matter. All other inputs support that process. Thus, substitution is possible across energy goods and prime movers, yet there are no substitutes for energy and power themselves. Higher efficiency can reduce the magnitude of energy transfers required to yield a given output, but there always exists a physical minimum (e.g. stoichiometric requirement in ammonia synthesis).

Accordingly, growth is the increase in the capacity to transfer energy, and can be studied with its constraints. The first one is the "energy budget constraint", which restricts the goods to be consumed by an agent such that the total energy transferred in their production is no greater than the agent's energy income. In the case of a simple autarkic agent, this constraint is

$$\boldsymbol{\gamma}_N^{A\prime}\boldsymbol{Q}_N + \boldsymbol{\gamma}_\epsilon^{A\prime}\boldsymbol{Q}_\epsilon = G \leq I = \boldsymbol{\delta}'_\epsilon \boldsymbol{Q}_\epsilon, \tag{1}$$

---

[1] In the case of a mechanical change, the second principle is a natural implication of Newton's second law, as material rearrangements require a change in acceleration of particles and therefore a transfer of energy.

where $\boldsymbol{Q}_N = [Q_1, \ldots, Q_N]'$ and $\boldsymbol{Q}_\epsilon = [Q_1, \ldots, Q_\epsilon]'$ are the quantities of non-energy and energy goods to be consumed, $\boldsymbol{\gamma}_N^A$ and $\boldsymbol{\gamma}_\epsilon^A$ are their respective average embodied energies, and $G$ is the total energy transferred. Average embodied energy is the total (direct plus indirect) energy transferred on average to produce a unit of a good.[2] $\boldsymbol{\delta}_\epsilon = [\delta_1, \ldots, \delta_\epsilon]'$ are energy good's energy contents which stem from Primary Energy Sources (PES), and $I$ is the agent's energy income.

This constraint must hold regardless if the agent is aware of it because energy must be available to be transferred, and therefore can be used to model an autarkic agent's behavior. Focusing on a single period, a world without uncertainty, and assuming the agent's preferences can be represented by a quasiconcave, continuous, and twice-differentiable utility function that is strictly increasing on non-energy goods, the agent's primary objective is to maximize

$$U = U(\boldsymbol{Q}_N), \tag{2}$$

subject to the energy budget constraint in (1). By rearranging such constraint as

$$\boldsymbol{\gamma}_N^{A\prime} \boldsymbol{Q}_N \leq E, \tag{3}$$

the FOCs resulting from maximizing (2) subject to (3) choosing quantities are

$$U_n/\lambda = \gamma_n, \quad \forall\, n \in N \tag{4}$$

where $E = \boldsymbol{\delta}'_\epsilon \boldsymbol{Q}_\epsilon - \boldsymbol{\gamma}_\epsilon^{A\prime} \boldsymbol{Q}_\epsilon$ is the agent's energy surplus, $U_n$ is good $n$'s marginal utility, $\lambda$ is the marginal utility of energy, $\gamma_n = \gamma_n^A(1 + \eta_n)$ is good $n$'s marginal embodied energy, and $\eta_n = \frac{\partial \gamma_n^A}{\partial Q_n} \frac{Q_n}{\gamma_n^A}$ is the quantity elasticity of average embodied energy. A good's marginal embodied energy is the total energy transferred to produce one more unit. The equilibrium condition in (4) states that the Marginal Rate of Substitution (MRS) between good $n$ and energy equals the good's marginal embodied energy. This does not mean that energy provides utility *per se*, but that more utility can be derived from one more joule due to the additional consumption it enables of utility-yielding non-energy goods.

Equation (4) can be used to derive tangency conditions between any two goods, which together with the energy budget constraint yield energetic-Marshallian demand functions $Q_n^*(\boldsymbol{\gamma}_N, E)$ and the marginal utility of energy surplus $\lambda^*(\boldsymbol{\gamma}_N, E)$. Replacing the demand functions back in (4) yields the unambiguous proposition that (a) if the marginal embodied energy of a non-energy good increases, *ceteris paribus*, its optimal level of consumption will fall, and that (b) if the marginal embodied energy of a non-energy good increases, *ceteris paribus*, the optimal level of another non-energy good will increase (details in Appendix A).

---

[2] For discussions on the concept of embodied energy see IFIAS (1974), Chapman (1974), Bullard & Herendeen (1975), Costanza (1981), and Huettner (1982). Here we overlook the issues of system boundaries and co-products, and assume that all energy transfers can be uniquely allocated between all goods.

## 2. Energy surplus maximization

The energy budget constraint as expressed in (3) implies that for Pareto-efficiency, utility maximization requires energy surplus maximization. This secondary objective can be stated as

$$max_{Q_\epsilon} E = \sum_{e=1}^{\epsilon}[\delta_e Q_e - G_e], \quad (5)$$

where $Q_e$ is the quantity produced of energy good $e$ and $G_e$ is the total energy transferred to produce such quantity. However, this objective function is subject to a "prime mover constraint" of the form

$$\sum_{n=1}^{N} x_{l,n} + \sum_{e=1}^{\epsilon} x_{l,e} \leq \bar{x}_l, \quad \forall\, l \in L \quad (6)$$

where $x_{l,n}$ is the quantity of prime mover of type $l$ used in the production of non-energy good $n$ and $\bar{x}_l$ is the agent's endowment of that prime mover type. Thus, the constraint restricts the employment of prime mover $l$ to the agent's endowment. Energy surplus maximization is also subject to an "energy usability constraint"

$$\sum_{l=1}^{L} \varepsilon_l \sum_{n=1}^{N} x_{l,n} \geq E, \quad (7)$$

where $\varepsilon_l = \int_t^{t+1} p_l\, dt$ is the direct energy transferred by each unit of prime mover $l$ during the period under study given its power rate $p_l$, which we assume constant for all prime movers of the same type. This constraint ensures that the agent does not produce useless energy surplus by forcing all energy surplus to be usable by prime movers employed in the production of utility-yielding non-energy goods. The energy usability constraint highlights that energy does not provide utility *per se*, but only through the capacity to produce other goods. This sets a distance with energy theories of value as suggested in Hannon (1973) and Costanza (1980), while simultaneously recognizing the role of energy in production.

Merging the constraints in (6) and (7) yields a Lagrangian of the form

$$\mathcal{L}_e = (1-\varphi)\sum_{e=1}^{\epsilon}[\delta_e Q_e - G_e] + \varphi \sum_{l=1}^{L} \varepsilon_l \bar{x}_l - \varphi \sum_{l=1}^{L} \varepsilon_l \sum_{e=1}^{\epsilon} g_{-l.e}(Q_e, \boldsymbol{x}_{-l,e}), \quad (8)$$

where $\varphi \in [0,1)$ is the proportion of useless energy surplus to unconstrained maximum energy surplus, which is a measure of aggregate prime mover scarcity ($\varphi = 0$ implies that all prime movers are readily available). Also, $g_{-l,e}(Q_e, \boldsymbol{x}_{-l,e})$ is the technical requirements of production function and $\boldsymbol{x}_{-l,e}$ are all prime movers used in the production of good $e$ apart from $l$. The FOCs choosing quantities are

$$\delta_e = \gamma_e + \alpha_e, \quad \forall\, e = 1 \ldots \epsilon \tag{9}$$

where $\alpha_e = \frac{\varphi}{1-\varphi} \frac{\sum_{l=1}^{L} \varepsilon_l g'_{-l,e}}{L}$ is energy good $e$'s marginal energy surplus, $g'_{-l,e}$ is the marginal technical requirements of production function, and $L$ is the quantity of prime mover types. This expression relates to the Energy Return Over Investment (EROI) literature (see Hall (2017)) as a good's marginal energy surplus is directly related to its marginal EROI (mEROI).[3]

Equation (9) also shows that whenever $\alpha_e > 0$, the agent optimally leaves energy surplus "on the table" despite such surplus being the explicit constraint for utility maximization.[4] This happens because of the underlying prime movers constraints that makes additional energy surplus useless. Thus, the equimarginal principle $\delta_e = \gamma_e$ only holds under no prime mover constraints.

The integral of $\alpha_e$ over the optimal production range for all energy goods yields the maximum useful energy surplus, such that if the optimal supply of good $e$ is $Q_e^*$ (computed using (9) and $x_l^*$ found below), the optimal solution to (5) is

$$E^* = \sum_{e=1}^{\epsilon} \int_0^{Q_e^*} \alpha_e \, dQ_e. \tag{10}$$

Figure 1 represents the equilibrium conditions specified in (9) and the solution obtained in (10). The black line is the agent's maximum willingness to transfer energy to produce one more unit, which is horizontal at the good's energy content up to the quantity that saturates the agent's aggregate power level $P = \sum_{l=1}^{L} p_l x_l$, and then falls vertically as the good's energy becomes useless. The Marginal Embodied Energy Curve (MEEC) indicates the marginal embodied energy at each level of production, which can be downward-sloping at some intervals due to efficiency gains from increasing returns to scale, yet will eventually be upward-sloping as the agent is forced to tap increasingly inconvenient PES. This dynamic has been documented for the production of natural resources since Ricardo and Malthus (Backhouse, 2004), and has been specifically modeled for energy goods in Court & Fizaine (2017) and Dale et al. (2011). The intersection between these two curves yields the optimal production $Q_e^*$ at marginal embodied energy $\gamma_e$, total energy transfer $G_e^*$, and energy surplus $E_e^*$. The marginal energy surplus $\alpha_e$ is the vertical distance between $\delta_e$ and $\gamma_e$.

---

[3] Dividing (9) by $\gamma_e$ yields $mEROI_e = 1 + \alpha_e/\gamma_e$, where $mEROI_e = \delta_e/\gamma_e$.
[4] $\alpha_e$ can be positive and comply with the First Law of Thermodynamics because $\delta_e$ comes from PES.

The logic behind the energy budget constraint in (3) should now be clear. The agent's energy income covers the energy transfers required to produce such income, and only the excess can be used to produce non-energy goods.[5] Accordingly, the optimal production of energy goods is independent from the agent's preferences. This result is set with the assumption that energy goods are not part of the utility function, yet its relevance rests with the energy budget constraint's logic. On the other hand, non-energy goods' optimal production depends on the agent's preferences. How energy is allocated to produce them is shown in Appendix B.

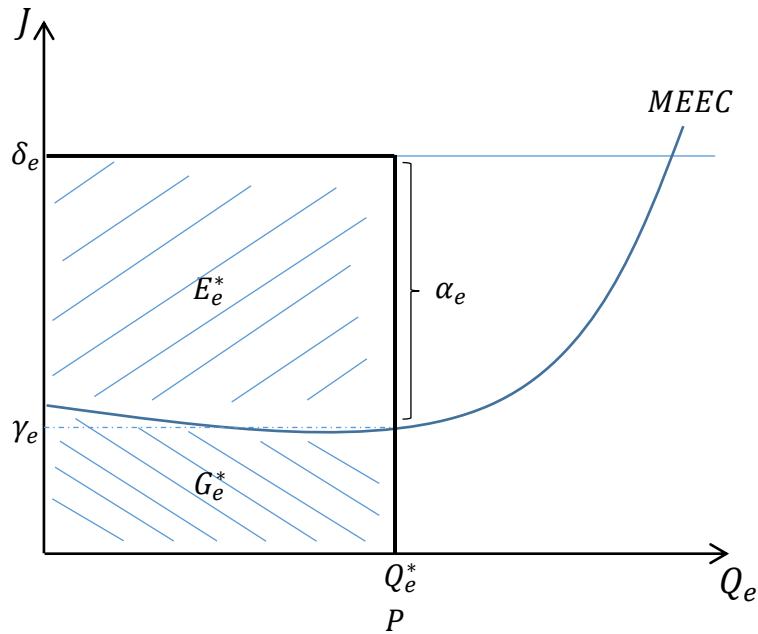

Figure 1. Equilibrium conditions in the production of an energy good

---

[5] Although the dualism embedded in modeling energy sectors underpinning the existence of non-energy sectors broadly recalls Physiocracy and Fei & Ranis (1963) model of development, the idea is closer to the model in Fizaine & Court (2016).

The Lagrangian in (8) can also be specified using prime movers as $Q_e = f_e(\boldsymbol{x_e})$, where $f_e(\cdot)$ is a concave-from-above, continuous, and twice-differentiable production function, and $G_e = \sum_{l=1}^{L} \omega_l x_{l,e}$, where $x_{l,e}$ is the quantity of prime mover $l$ used in the production of the good and $\omega_l = \varepsilon_l + d_l \gamma_l^A$ is the total energy transferred by each unit of the prime mover. Note that $\omega_l$ contains direct energy transfers $\varepsilon_l$ and indirect ones that depend on the prime mover's average embodied energy $\gamma_l^A$ and depreciation rate $d_l \in (0,1)$. Solving the Lagrangian with this alternative specification yields FOCs choosing prime movers as

$$\delta_e f_{l,e} = \omega_l + \phi_l, \qquad \forall\, l = 1 \in L \qquad (11)$$

where $f_{l,e}$ is prime mover $l$'s marginal productivity in the production of good $e$ and $\phi_l = \frac{\varphi \varepsilon_l}{1-\varphi}$ is prime mover $l$'s marginal energy surplus. Equation (11) implies that a prime mover's marginal energy income covers the prime mover's total energy transfers and leaves a surplus, which is key to growth dynamics as shown in the next section.

Equation (11) can be used to derive tangency conditions that specify optimal production conditions for any two prime movers and any two energy goods. The former conditions depend on the prime movers' marginal productivities on the same good and total energy transfers, while the latter conditions depend on each prime mover's marginal productivity on different energy goods and such goods' energy contents. Also, replacing back the optimal prime mover derived demands in (11) yields the unambiguous proposition that (c) if the energy content of an energy good increases, *ceteris paribus*, its optimal level of production increases (details in Appendix C).

### 3. Growth

Growth is driven by positive marginal energy surplus of energy goods, yet materializes with prime mover accumulation. Despite securing $\alpha_e$ if $Q_e^*$ is increased by one unit, such marginal surplus is useless to the agent due to prime mover constraints and therefore does not lead to greater production by itself. Yet, $\alpha_e > 0 \Rightarrow \phi_l > 0$ because $\alpha_e = \frac{1}{L}\sum_{l=1}^{L} \phi_l g'_{-l,e}$. If $\phi_l > 0$, the agent has incentives to produce more prime movers for the following period, which relaxes its prime mover constraint, and leads to higher production of energy goods, securement of energy surplus, production of non-energy goods, and therefore to higher utility.

Formally modeling prime mover accumulation requires at least a two-period model that surpasses the scope of this paper. Nonetheless, such accumulation can be conceptualized with a modified logistic model used for population growth of the form

$$\frac{dx_l}{dt} = r_l tanh(\phi_l) x_l, \qquad \forall\, l = 1 \in L \tag{12}$$

where $\frac{dx_l}{dt}$ is the increase of prime mover of type $l$, $r_l$ is its maximum unitary rate of accumulation, and $tanh$ is the hyperbolic tangent function. Equation (11) implies that under large $\phi_l$ the $tanh \rightarrow 1$, and the prime mover type accumulates at an unconstrained rate. Yet, *ceteris paribus*, as the prime mover accumulates, $\phi_l$ and $tanh$ fall such that accumulation slows down. Under no prime mover constraints, $\phi_l = 0$ and $tanh(0) = 0$ such that accumulation stops.

Figure 2 represents the increase in the production of energy good $e$ from accumulating prime movers. As in Figure 1, at $t = 0$ demand is perfectly elastic at $\delta_e$ up to the quantity that saturates the agent's aggregate power level $P^0$, where it falls vertically. As prime movers accumulate given $tanh(\phi_l) > 0\ \forall\, l$, the prime mover constraints are relaxed such that at $t = 1$ the agent produces $Q_e^{1*}$ and additionally secures $E^{1*}$. This increases the agent's utility, yet the good's marginal energy surplus is even higher and therefore $tanh(\phi_l) > 0$. At $t = 2$, a higher aggregate power level leads to the production of $Q_e^{2*}$ where the good's marginal embodied energy increases to $\gamma_e^2$, but this increase is small enough for $\alpha_e > 0$. Finally, at $t = 3$ the production of $Q_e^{3*}$ implies $\alpha_e = 0$ and $tanh(0) = 0$. In brief, when energy goods are relatively abundant (i.e. their marginal embodied energy and marginal EROI is high) prime movers place the constraint on growth, yet when energy goods are scarce they become the constraint (Fizaine & Court, 2016b; Stern, 2011).

The agent's long-run equilibrium or steady-state $\zeta$ is guaranteed to exist if the MEEC is at least eventually upward-sloping.[6] In $\zeta$ no more prime movers can be accumulated, and therefore the agent can be said to have maximized its aggregated power level as suggested in Lotka (1925) and Odum (1995). Also, in $\zeta$ no additional energy surplus can be harnessed given the energy flows governing production, and therefore utility cannot be further increased by the mechanisms discussed up to here.

---

[6] Long-run economic equilibrium is a thermodynamic disequilibrium (Prigogine 1996), which implies that economies are systems that evolve towards states far from thermodynamic equilibrium (Ayres, 1998).

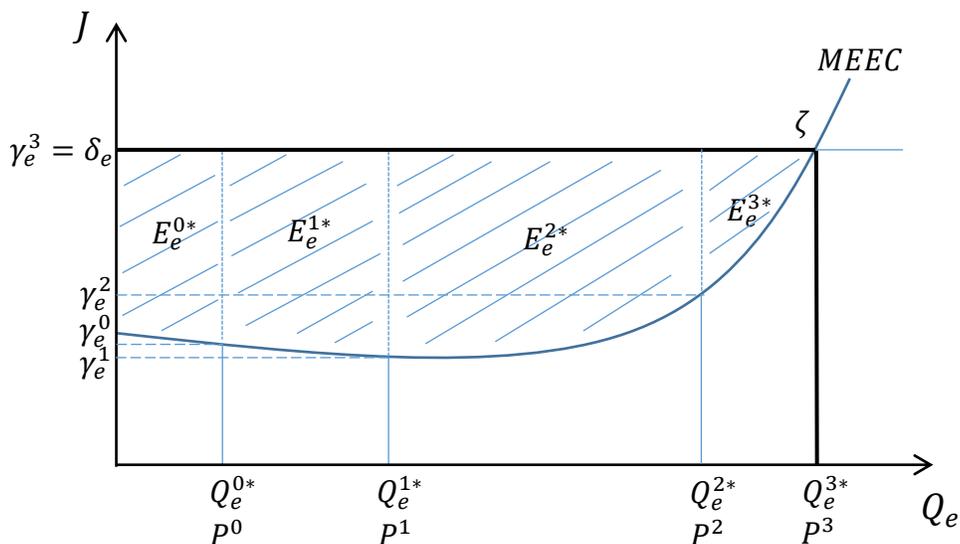

Figure 2. Growth dynamics driven by energy good's marginal energy surplus

Yet, there are secondary components of energy transfers that move $\zeta$ constantly. Efficiency enhancements shift the MEEC downward and push $\zeta$ to the right (see Appendix D for a discussion on efficiency), depletion of Primary Energy Sources (for non-renewable energy goods) shifts the MEEC upwards and push $\zeta$ to the left, and changing weather and security conditions affect the MEEC and can move $\zeta$ in either direction. Also, the invention or discovery of new prime movers (e.g. steam engines) and energy goods (e.g. electricity) can dramatically shift existing MEECs, and by creating entirely new ones, lead to long new growth spells. This is the underlying narrative in the historical accounts in White (1943), Cook (1976), Smil (1994), and Weissenbacher (2009).

## 4. Conclusion

Our results show that energy goods' positive marginal energy surplus drives growth by incentivizing prime mover accumulation and enabling higher utility levels. Growth processes cannot be understood independently from the energy goods energizing them. Yet, the relevance of such goods must be stated in relation to the prime movers transferring their energy contents and agents' desires to rearrange matter. The focus on energy transfers using canonical economic methodology offers an alternative to conventional growth theories that recognizes energy's prominence as suggested by natural science, while avoiding energy determinism. This framework, when applied to a setting with exchange, markets, and prices, provides a theoretical explanation of the observed proportionality between market prices and embodied energies (Costanza, 1980; Gutowski et al., 2013; Liu et al., 2008). Monetary values as social expressions of underlying energy dynamics suggest a general reinterpretation of economics as an interplay between human desires and thermodynamic processes.